\documentclass{article}
\usepackage[utf8]{inputenc}
\usepackage{amsmath,amsfonts}
\usepackage{algorithmic}
\usepackage{algorithm}
\usepackage{array}
\usepackage[caption=false,font=normalsize,labelfont=sf,textfont=sf]{subfig}
\usepackage{textcomp}
\usepackage{stfloats}
\usepackage{url}
\usepackage{verbatim}
\usepackage{graphicx}
\usepackage{xcolor}
\usepackage{cite}
\usepackage{tikz}
\usetikzlibrary{quantikz} 
\usepackage{enumitem}
\usepackage{authblk}

\usepackage[toc,page]{appendix}

\usepackage{lineno}

\title{Federated Learning with Quantum Secure Aggregation}
\author[1]{Yichi Zhang}
\author[1]{Chao Zhang}
\author[2]{Cai Zhang}
\author[1]{Bei Zeng}
\author[3]{Qiang Yang}
\author[4]{Lixin Fan\footnote{lixin.fan01@gmail.com}}
\affil[1]{Department of Physics, The Hong Kong University of Science and Technology}
\affil[2]{College of Mathematics and Informatics, South China Agricultural University}
\affil[3]{Department of Computer Science and Engineering, The Hong Kong University of Science and Technology}
\affil[4]{WeBank}

\begin{document}
% \linenumbers
\maketitle

\abstract{
This article illustrates a novel Quantum Secure Aggregation (QSA) scheme that is designed to provide highly secure and efficient aggregation of local model parameters for \textit{federated learning}. The scheme is \textit{secure} in protecting private model parameters from being disclosed to semi-honest attackers by utilizing quantum bits i.e. \textit{qubits} to represent model parameters. The proposed security mechanism ensures that any attempts to eavesdrop
private model parameters can be immediately detected and stopped. The scheme is also \textit{efficient} in terms of the low computational complexity of transmitting and aggregating model parameters through entangled qubits. 

Benefits of the proposed QSA scheme are showcased in a horizontal federated learning setting in which both a \textit{centralized} and \textit{decentralized} architectures are taken into account. It was empirically demonstrated that the proposed QSA can be readily applied to aggregate different types of local models including logistic regression (LR), convolutional neural networks (CNN) as well as quantum neural network (QNN), indicating the \textit{versatility} of the QSA scheme. Performances of global models are improved to various extents with respect to local models obtained by individual participants, while no private model parameters are disclosed to \textit{semi-honest} adversaries.
}

\section{Introduction}

In the era of big data with trillion bytes of raw data being distributed across devices or institutions, Federated Learning (FL) has been introduced as an effective technology to allow multiple parties to collaboratively train a machine learning model \textit{without gathering or exchanging private training data} and complying with regulatory requirements such as General Data Protection Regulation (GDPR)\footnote{GDPR is applicable as of May 25th, 2018 in all European member states to harmonize data privacy laws across Europe. https://gdpr.eu/}.  A crux of the matter in Federated Learning is the \textit{secure aggregation} of local models that are firstly trained by multiple parties and subsequently aggregated for the sake of improved aggregated model performances. The aggregation is required to be {secure} in the face of \textbf{semi-honest} attacks which aim to stealthy spy private data that belong to respective parties.  On the other hand, the aggregation has to be computationally and communicationally efficient to be useful for practical applications.  How to ensure that the aggregation is both \textit{secure} and \textit{efficient} without compromising performances of the aggregated model is the key challenge investigated in our recent work by utilizing \textit{quantum computation and communication}.  

For FL with \textit{classical computations}, the mission of secure model aggregation is often achieved by adopting privacy computing techniques such as Homomorphic Encryption (HE) \cite{PHE, FHE} or Differential Privacy (DP) \cite{DP1, DP2}.  HE is considered to be secure against semi-honest attacks with provable security guarantees \cite{DP3, DP4}.  Yet HE approaches incur significant computational and communication overhead, often orders of magnitudes less efficient as compared with alternative privacy computing techniques \cite{HEsurvey}. High computational and communication complexity of HE makes it hardly applicable to protect large machine learning models e.g. deep neural networks with billions of parameters\footnote{By adopting some ad-hoc tricks like batch encryption and quantized  gradients etc., Zhang et. al. shown that one can train a \textit{medium-sized} neural network with 102K of gradient parameters protected by Paillier Homomorphic Encryption \cite{BatchCrypt} (see discussions in Sect. \ref{sect:disc}). }. Differential Privacy (DP), on the other hand, is much more efficient and readily applicable to complex DNN models \cite{DP2}. Yet it has been shown that under certain circumstances semi-honest attackers may exploit the information exposed for  model aggregation and recover private training data \cite{DP3, zhu_deep_2020}.  The secure aggregation with efficient classical computation remains to be an active research topic for both theoretical investigation \cite{NFL} and empirical study \cite{BatchCrypt}. 

Bearing in mind the aforementioned perplexity of model aggregation using classical computation, we demonstrate in this paper a fundamentally different and novel approach by exploiting a quantum communication protocol that leads to both \textit{secure} and \textit{efficient} FL model aggregation.  
In our view the proposed quantum computing based solution is compared favorably against its classical computing based counterparts, due to three distinctive features summarized below and illustrated in following sections of this article: 

\begin{itemize}

\item First, the GHZ aggregation scheme is highly \textit{efficient} since it does not incur any computationally demanding encryption to protect private model parameters. Instead, these parameters are  efficiently encoded and protected in quantum state (see Sect. \ref{Sect:QSA}). 

\item Second, the proposed GHZ aggregation  is \textit{secure} since it adopts a quantum secret sharing protocol \cite{QSS} to aggregate local models trained by multiple federated learning participants. The privacy of training data and local models are guaranteed to be protected by the unique physical properties of quantum communication, with which potential eavesdropping can be immediately detected and prevented (see analysis in Sect. \ref{Sect:Security}). 

\item Third, as demonstrated in Sect. \ref{Sect:Experiment}, the GHZ aggregation is also \textit{versatile} in the sense that it can be readily used to protect either classical Federated Learning models (e.g. a neural network model) or a quantum neural network (QNN) model which allows a large set of batched trained data to be processed in parallel by quantum circuits in QNN. 

\end{itemize}

The rest of the present article is organized as follows. Sect. \ref{sect:related} briefly reviews existing work related to federated learning using some forms of quantum computing or quantum communications. However, none of these work aim to simultaneously improve \textit{security, efficiency} and model \textit{performances} as presented in our work. For the sake of completeness, Sect. \ref{sect:preliminary} presents necessary background information about federated learning and quantum computing \& communication. Sect. \ref{sect:QFL} illustrates overall architectures of the proposed QFL framework followed by provable security guarantee of the quantum aggregation. Efficiency and superiority brought by quantum communication  and computing are also illustrated. In order to demonstrate applicability of the proposed framework, Sect. \ref{Sect:Experiment} illustrates noticeable improvements in the federated model performances for a variety of horizontal federated learning settings including both I.I.D and non-I.I.D data distributions\footnote{We refer readers to Appendix \ref{Sect:QNN} for detailed illustration of experimental results and explanations about the simulator of QNN networks.}.
\subsection{Related Work}\label{sect:related}

It was shown that Quantum computing can be used for machine learning to reduce computational complexity and improve model performances as demonstrated for principle component analysis, support vector machines and neural networks etc. \cite{QML-2013,QSVM-2014,QML_survey_Nat17,QNNspeedup_1,QNNspeedup_2,QNNspeedup_3}. Recent research efforts alone this line have also been devoted to \textit{federated learning} under the name of \textit{quantum federated learning} e.g. in \cite{QFL_related_work_1, QFL_related_work_2, QFL_related_work_3, QFL_related_work_4, QFL_related_work_5, du2021learnability, du2022efficient, du2022quantum}. However, these work focused on boosting federated learning model performances by leveraging the computational power of quantum computers, and employed classical measures e.g. Homomorphic Encryption or Differential Privacy to provide security protection for federated learning against semi-honest adversaries. 

\textbf{Distributed quantum secure protocol} was proposed to protect exchanged information from eavesdropping for distributed machine learning in general \cite{sheng2017distributed}. This work is probably the most similar to our proposed QFL framework, and the application scenario involving multiple hospitals to jointly provide certain machine learning capabilities was well-motivated. Nevertheless, the machine learning task in \cite{sheng2017distributed} was over-simplified as calculating the distance between two two-dimensional vectors, and it is unclear how the calculating precision and efficiency were affected by their protocol since neither theoretical analysis nor experimental results were provided in \cite{sheng2017distributed}.  

Note that necessary background knowledge and related works about federated learning, differential privacy, homomorphic encryption, quantum computing and communication etc. are reviewed in the following section.

\section{Preliminary}\label{sect:preliminary}

This section briefly reviews  necessary background information about federated learning and quantum computing \& communication. For readers with a variety of backgrounds in computer science or physics, this review aims to provide a self-contained introduction about motivations, architectures and requirements of federated learning, as well as fundamentals of quantum computing and communication from a perspective of machine learning or information processing.

\subsection{Federated Learning}
Federated Learning (FL) is a suite of distributed machine learning technology that aims to collaboratively develop a machine learning model from distributed data ownered by respective parties while protecting privacy of such data.
The term \textit{federated learning} was first coined by McMahan et al. \cite{mcmahan2016federated,mcmahan2017communication, konevcny2016federated, konevcny2016federated_new}, which demonstrated how to aggregate local models learned on multiple devices for the sake of improved model performances, yet, without sending private data to a \textit{semi-honest} third-party server or other devices. Bearing in mind the privacy concern of secret data distributed across multiple institutions, Yang et al. \cite{yang2019federated} extended Federated Learning application scenarios into three categories: 
\begin{itemize}
    \item \textit{Horizontal federated learning}, in which datasets share the same feature space but different space in samples; 
    \item \textit{vertical federated learning} in which  two datasets share the same sample ID space but differ in feature space; 
    \item \textit{Federated transfer learning} in which two datasets differ not only in samples but also in feature space.
\end{itemize}

In this article we only focus on horizontal federated learning with a centralized and decentralized architectures (see Fig. \ref{QFL_structure}). We refer readers to \cite{mcmahan2016federated,mcmahan2017communication, konevcny2016federated,konevcny2016federated_new,yang2019federated,DBLP:journals/ftml/KairouzMABBBBCC21} for detailed and thorough treatments of federated learning.

\subsubsection{Threat model} \label{subsect:threat}
\begin{figure}
    \centering
    \includegraphics[width=\linewidth]{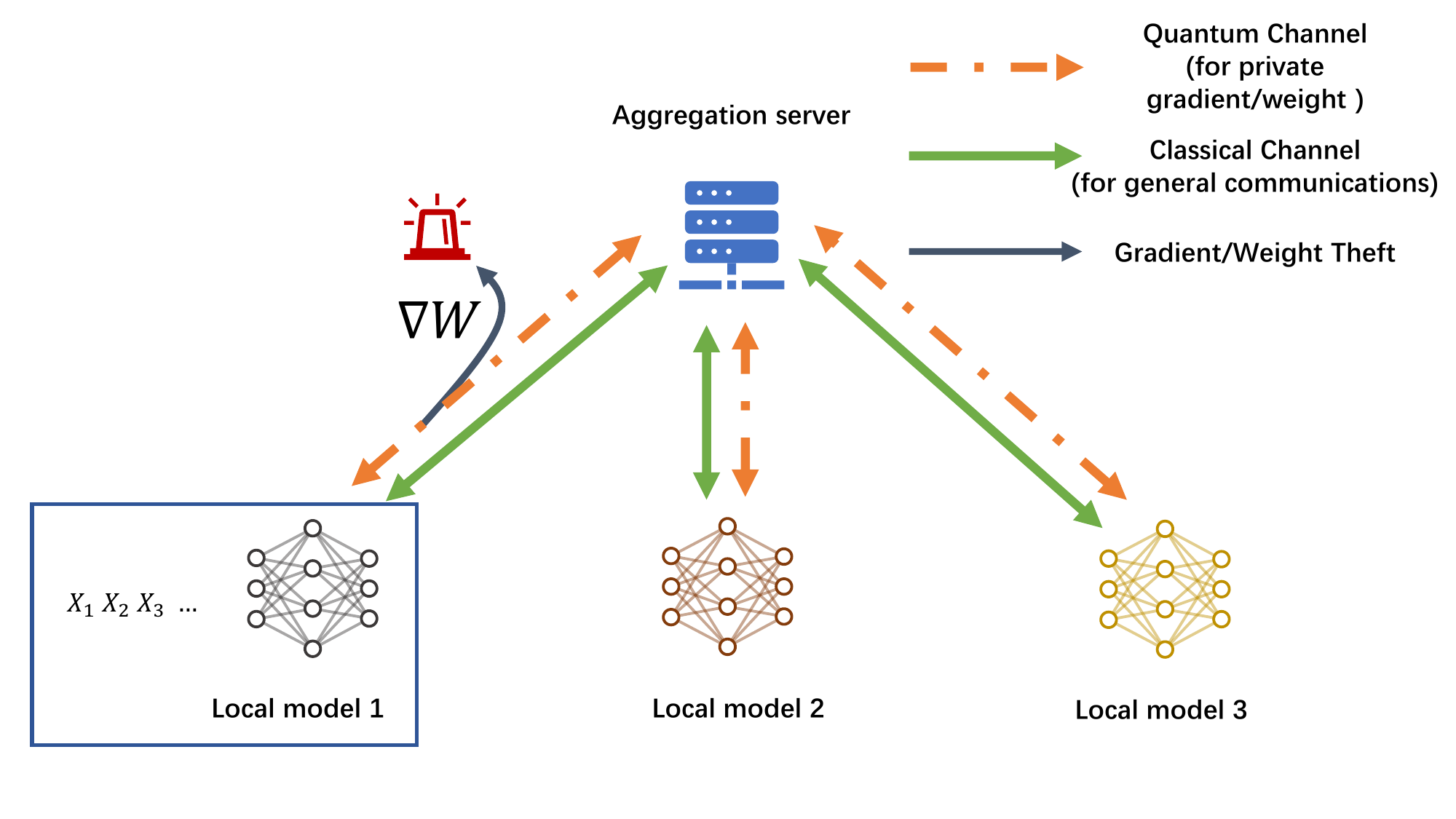}
    \caption{Bayesian Inference Attacks aim to infer private training data from the exchanged information in the learning process e.g. model gradients $\nabla W$ that are not properly protected. 
    }
    \label{fig:DLG-attack}
\end{figure}

It is often assumed that in a horizontal federated learning setting  there exist one or more \textit{semi-honest} adversaries, who, on one hand, adhere to the protocols regulating legitimate federated learning behaviors of each party, and on the other hand, attempt to stealthy spy private data that belong to other parties. That is to say, semi-honest adversaries appear perfectly normal in terms of their public behaviors. For instance, one recent report showcased that adversaries can reconstruct other parties' private training images up to pixel-level accuracy, by adopting a Bayesian Inference attack on information exchanged during the learning process e.g. mode gradients that are not properly protected \cite{zhu_deep_2020,NFL}. In particular, Fig. \ref{fig:DLG-attack} illustrates a scheme of such an attack in which adversaries build up the reconstruction pipeline on his/her own private machines completely outside the normal federated learning process. It is therefore impossible to detect such misbehavior of semi-honest adversaries who appear indistinguishable from benign parties.  
For federated learning employing quantum secure aggregation, however, semi-honest parties can no longer hide its spying behaviors due to the unique physical properties of quantum communication, and such misbehaves in QFL can be caught and prevented on spot (see Sect. \ref{Sect:Security} for details).  

\subsubsection{Privacy protection mechanisms} 

For classical computing, the protection of private information is often achieved by either ``smearing'' information to be protected with random perturbations or ``hiding'' secret information with carefully designed transformations of secrets. The well-established \textit{differential privacy} (DP) technique belongs to the former approaches \cite{dwork2006differential,dwork2014algorithmic,DP1,DP2,DP3,DP4}, while the influential \textit{homomorphic encryption} (HE) is representative to the latter approaches \cite{FHE,PHE,HEsurvey}. For the sake of completeness of this article, we briefly summarize respective principles, merits and shortcomings of both DP and HE below. 
\begin{itemize}%[left=0]
    \item Differential privacy (DP) is a celebrated privacy-preserving mechanism initially proposed by Dwork et al. \cite{dwork2006differential,dwork2014algorithmic,DP1} to protect individual privacy by adding random perturbations in response to queries about databases. Formally, \textit{a randomized mechanism $\mathcal{M} : \mathcal{X} \rightarrow \mathcal{R}$ with domain $\mathcal{X}$ and range $\mathcal{R}$ is ($\epsilon$,$\delta$)-differentially private, if for all measurable sets $\mathcal{S} \subseteq \mathcal{R}$ and for any two adjacent databases $D_i,D'_i \in \mathcal{X}$ \cite{dwork2014algorithmic}, 
    \begin{align}\label{eq:DPdef}
      \Pr[ \mathcal{M}(D_i) 
      \in S] \leq e^{\epsilon} \Pr[ \mathcal{M}(D'_i) \in S]+ \delta.
    \end{align}}
    In essence, the definition (\ref{eq:DPdef}) dedicates that the protection mechanism $\mathcal{M}$ makes two private databases indistinguishable up to a constant $e^\epsilon$ close to 1.  
    DP was later widely adopted to protect privacy for federated learning methods including deep neural networks \cite{abadi2016deep} and other machine learning models \cite{DP1}.

    \item Homomorphic encryption (HE) includes fully homomorphic encryption (FHE) \cite{FHE} supporting arbitrary operations and partially homomorphic encryption (PHE) \cite{PHE} supporting limited operations. Homomorphic encryption can host data computing tasks to third party under the premise of ensuring privacy and accuracy. A brief HE process can be described as following:\\
    $pk, sk\leftarrow Key.Gen(1^\lambda)$: The participant generates a set of public key $pk$ and secret key $sk$ locally. \\
    $c \leftarrow Enc_{pk}(m)$: The participant encrypts the data $m$ locally with the public key and obtains the ciphertext $c$. \\
    $Enc_{pk}[f(m_1, m_2, ...)] \leftarrow Eval(f, c_1, c_2,...)$: The server calculates the ciphertext $c_1, c_2,...$ with modified function $f$ according to the convention $Eval$. The result obtained is the same as the result of directly calculating the plaintext $m_1, m_2, ...$ with function $f$ and then encrypting it.\\
    $m \leftarrow Dec_{sk}(c)$: The participant decrypt the result locally.\\
    Although homomorphic encryption can guarantee the security of computation, its huge computational overhead makes it difficult to be practical.
\end{itemize}

The secure aggregation with efficient classical computation remains an active research topic. In particular, Zhang et. al. \cite{NFL} treated the optimal privacy-utility trade-off as a constrained optimization of \textit{utility loss} $\epsilon_u$ subject to a given bound on privacy leakage $\epsilon_p$ as follows,

A theoretical analysis of the trade-off was then manifested as the \textit{No Free Lunch Theorem} which dictates that, when DP like protection mechanisms are adopted, one has to trade a decrease of the privacy-loss with a certain degree of increase of the utility loss and vice versa. It was also shown that when  fully homomorphic encryption (FHE) or Paillier schemes are adopted, one has to trade security with substantially more demanding computational and communications costs. 

Moreover, this dilemma has also been addressed by adopting some ad-hoc tricks like batch encryption and quantized  gradients. For instance, Zhang et. al. \cite{BatchCrypt} shown that with reasonable amount of resource i.e. in hundreds of seconds and hundreds of MB of memory footprint, one can successfully train a \textit{medium-sized} neural network with 102K of gradient parameters protected by Paillier Homomorphic Encryption  (see discussions in Sect. \ref{sect:disc}).

\subsection{Quantum Computing}

\subsubsection{Qubit and Quantum State}

\textit{Qubit} in quantum computing is similar to \textit{bit} in classical computing. Qubit can be in logical value 0 and 1 like classical bit, usually denoted as the state $ 0\to |0\rangle $ and $1\to |1\rangle $ in Dirac notations. Besides, qubit can stay in the superposition of both states
$ | \psi\rangle=a|0\rangle+b|1\rangle$
where the complex number $a (b)$, also called probability amplitude, represents the state $|\psi\rangle$ can be found in state $0 (1)$ with probability $|a|^2 (|b|^2)$. The states $|0\rangle$ and $|1\rangle$ serve as the orthogonal basis in Hilbert space. Usually the basis $|0\rangle$ and $|1\rangle$ can be omitted without ambiguity. In this way the state can also be written as a column vector
$|\psi\rangle = [a,b]^T$
where $T$ is the matrix transpose. The normalization requirement for states
$ \langle \psi|\psi\rangle=|a|^2+|b|^2=1$
also means that the summation of all probabilities is one. Two isolated qubits can be described using Kronecker product
\begin{align*}
  |\psi_0\rangle \otimes |\psi_1\rangle &= (a_0|0\rangle+b_0|1\rangle) \otimes (a_1|0\rangle+b_1|1\rangle)\\
  &= a_0a_1 |00\rangle + a_0b_1 |01\rangle + b_0a_1 |10\rangle + b_0b_1 |11\rangle
\end{align*}
where the complex coefficients satisfy
$ |a_0|^2+|b_0|^2=1, |a_1|^2+|b_1|^2=1 $
More general and interesting states for two qubits are entangled,
$ |\psi\rangle = c_0 |00\rangle + c_1 |01\rangle + c_2 |10\rangle + c_3|11\rangle$
which usually cannot be written as the Kronecker product form. The normalization condition is
$ |c_0|^2+|c_1|^2+|c_2|^2+|c_3|^2=1$.
When the orthogonal basis $|00\rangle |01\rangle |10\rangle |11\rangle$, also called computational basis, are implicitly omitted as usual, the state can be written as
$ |\psi\rangle = [c_0,c_1,c_2,c_3]^T$
By induction, a $n$-qubit state is a unit normalized vector of length $2^n$ in Hilbert space. Some frequently-seen qubits states are summarized in table.

\begin{table}[h!]
    \centering
    \begin{tabular}{c|c}
        \hline
        name & vector representation \\
        \hline
        $|0\rangle$ & $[1,0]^T$\\
        $|1\rangle$ & $[0,1]^T$\\
        $|+\rangle$ & $\frac{1}{\sqrt{2}}[1,1]^T$\\
        $|-\rangle$ & $\frac{1}{\sqrt{2}}[1,-1]^T$\\
        Bell state $\frac{1}{\sqrt{2}}(|00\rangle + |11\rangle)$ & $\frac{1}{\sqrt{2}}[1,0,0,1]^T$\\
        GHZ state $\frac{1}{\sqrt{2}}(|000\rangle + |111\rangle)$ & $\frac{1}{\sqrt{2}}[1,0,0,0,0,0,0,1]^T$\\
        \hline
    \end{tabular}
    \caption{Frequently-seen qubit states}
    \label{table:qubit-state}
\end{table}

\subsubsection{Quantum Gates}
A quantum gate can be considered as a matrix, which can adapt on certain qubit(s). Applying quantum gate will transform qubits among different states. For example, Pauli-X gate, also called NOT gate, swaps the basis states $|0\rangle$ and $|1\rangle$,
$X|0\rangle=|1\rangle,\; X|1\rangle=|0\rangle $,
but leaves the states $|+\rangle$ and $|-\rangle$ unchanged
$X|+\rangle=|+\rangle,\; X|-\rangle=-|-\rangle $
where the global phase factor $e^{i\pi}=-1$ cannot distinguish the states $|-\rangle$ and $-|-\rangle$. For this reason, the states $|+\rangle$ and $|-\rangle$ are called the eigenstates of the Pauli-X gate. Quantum gates are linear operators and can be fully described as a unitary matrix $U$ satisfying
$ UU^\dag = U^\dag U = I$
where $\dag$ is the matrix conjugate transpose. The matrix representation for Pauli-X gate is
$ X=\bigl( \begin{smallmatrix}0 & 1\\ 1 & 0\end{smallmatrix}\bigr)$.
According to the number of operation qubits, quantum gates can be classified as $1$-qubit, $2$-qubit, or $n$-qubit gates. Apparently, Pauli-X gate belongs to $1$-qubit gates. An example of $2$-qubit gate is CNOT gate which flips the second qubit when the first qubit in the state $|1\rangle$. $1$-qubit and $2$-qubit gates from a set of universal quantum gates and can construct any $n$-qubit gate.

\begin{table}[h!]
    \centering
    \begin{tabular}
    { >{\centering\arraybackslash} m{1.5cm} | >{\centering\arraybackslash} m{4.5cm} | >{\centering\arraybackslash}m{1.5cm}}
        \hline
        name & matrix representation & symbol \\
        \hline
        Pauli-X & $\begin{bmatrix} 0&1\\1&0 \end{bmatrix}$ & \includegraphics[width=0.1\textwidth]{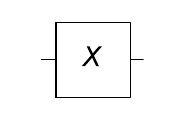} \\
        \hline
        Pauli-Z & $\begin{bmatrix} 1&0\\0&-1 \end{bmatrix}$ & \includegraphics[width=0.1\textwidth]{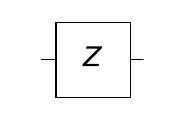} \\
        \hline
        Hadamard & $\frac{1}{\sqrt{2}}\begin{bmatrix} 1&1\\1&-1 \end{bmatrix}$ & \includegraphics[width=0.1\textwidth]{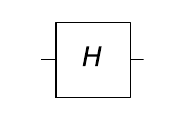} \\
        \hline
        $R_x$ & $\begin{bmatrix}
            \cos \frac{\omega}{2}&		-i\sin \frac{\omega}{2}\\
            -i\sin \frac{\omega}{2}&		\cos \frac{\omega}{2}\\
        \end{bmatrix}$ & \includegraphics[width=0.1\textwidth]{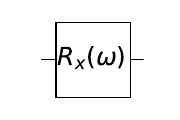} \\
        \hline
        $u_3$ & \scalebox{0.8}{${\begin{aligned}
            \hat{n}=\left( \sin \theta \cos \phi ,\sin \theta \sin \phi ,\cos \theta \right)\\
            u_3( \alpha ,\theta ,\phi) =I\cos \alpha +i\left( \hat{n}\cdot \vec{\sigma} \right) \sin \alpha
        \end{aligned}}$} & \includegraphics[width=0.1\textwidth]{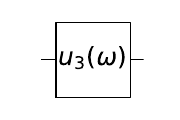} \\
        \hline
        CNOT & $\begin{bmatrix} 1&0&0&0\\0&1&0&0\\0&0&0&1\\0&0&1&0 \end{bmatrix}$ & \includegraphics[width=0.1\textwidth]{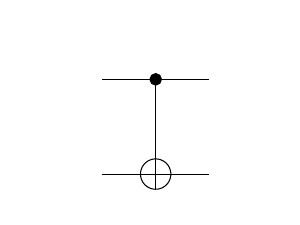} \\
        \hline
        C-$u_3$ & $\begin{bmatrix}
            I_{2\times 2}&		0_{2\times 2}\\
            0_{2\times 2}&		u_3\\
        \end{bmatrix}$ & \includegraphics[width=0.1\textwidth]{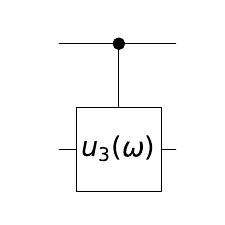}\\
        \hline
    \end{tabular}
    \caption{Frequently-seen quantum gates}
    \label{table:quantum-gate}
\end{table}

\subsubsection{GHZ State}

GHZ state \cite{GHZstate} is a typical quantum entangled state with simple structure and easy extension. This quantum state has the property of symmetry, and when any one qubit in the quantum state is measured, the rest of the qubits collapse to the same state. In addition, operating on any one of the qubits will affect the state of the entire system. These properties of GHZ state can be used to implement secure multi-party computation in quantum computing.

\subsubsection{Quantum Circuits}

\begin{figure}
    \centering
    \includegraphics[width=0.5\linewidth]{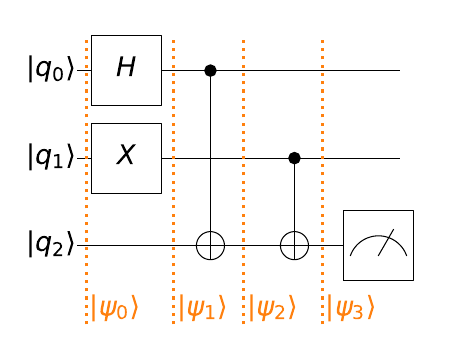}
    \caption{A Sample Quantum Circuit Diagram}
    \label{fig:simple-qnn}
\end{figure}

Similar to classical circuits, the quantum circuits model quantum computation and describe the quantum gates and measurements in sequence. A sample quantum circuit with $3$ qubits is plotted in Fig.\ref{fig:simple-qnn}. The horizontal lines represents the qubits $q_0,q_1,q_2$ separately. Reading from the leftmost of the figure, the quantum state is usually initialized to the basis state $|\psi_0\rangle=|000\rangle$. After applying the Hadamard gate on qubit $q_0$ and the Pauli-X gate on qubit $q_1$, the state turns into
$ |\psi_1\rangle=\frac{1}{\sqrt{2}}(|010\rangle + |110\rangle). $
Since these two quantum gates are on different qubits, they can be performed simultaneously. After the following two CNOT gates, the state becomes
$ |\psi_2\rangle=\frac{1}{\sqrt{2}}(|010\rangle + |111\rangle). $
$ |\psi_3\rangle=\frac{1}{\sqrt{2}}(|011\rangle + |110\rangle). $
Finally, measurements are made on the qubits we interested in. For this naive quantum circuit, we will get both measurement results $0$ and $1$, each with one half probability $P(0)=P(1)=0.5$.

\section{Federated Learning with  Quantum Secure Aggregation}\label{sect:QFL}

We illustrate in this section a novel quantum communication based approach that aims to boost both  \textit{security} and  \textit{efficiency} of Federated Learning (FL) without compromising \textit{performances} of the aggregated model in a horizontal Federated Learning (HFL) setting.
We first illustrate \textit{architectures} of the proposed Quantum Federated Learning (QFL) system which consists of both local models and the quantum secure aggregation (see Fig. \ref{QFL_structure} in Sect. \ref{subsect:architect}) Then we elaborate on the \textit{quantum secure aggregation protocol} adopted in the QFL (Sect. \ref{Sect:QSA}), in terms of its designing principle, time complexity as well as \textit{security analysis} (Sect. \ref{Sect:Security}).

\subsection{Architectures}\label{subsect:architect}
\begin{figure}
    \centering
    \includegraphics[scale=0.4]{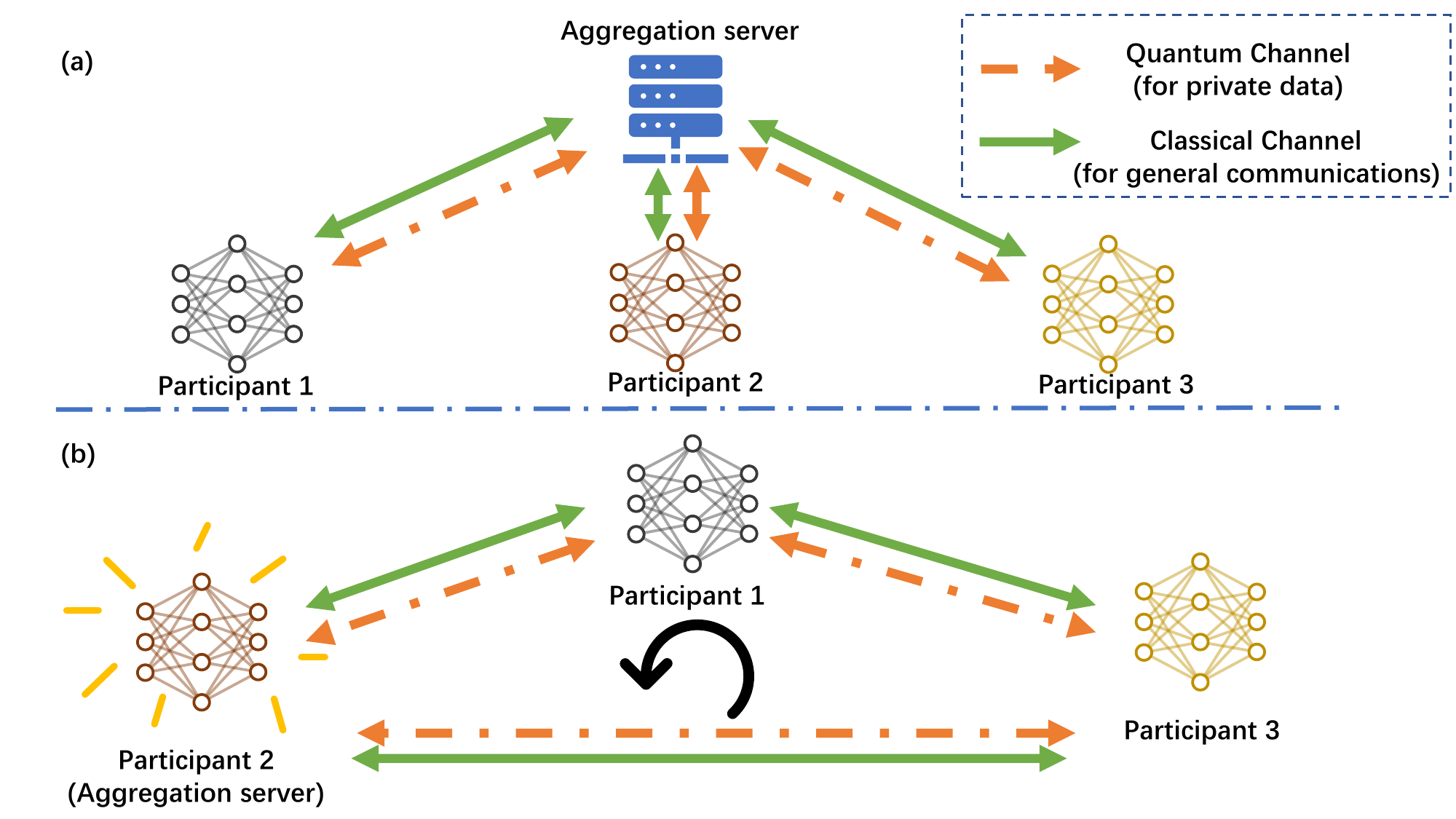}
    \caption{Two architectures of QFL: (a) with an aggregation server which sends via classical channels the aggregated model parameters to local parties ; (b) without  any aggregation server or classical channel for transferring of aggregated model parameters (see text in Sect. \ref{subsect:architect}).}
    \label{QFL_structure}
\end{figure}

Fig. \ref{QFL_structure} illustrates two different architectures, namely, a \textit{centralized} and a \textit{decentralized} structures of the proposed horizontal \textit{Quantum Federated Learning} (QFL) system, which consists of following parts as its building blocks.

\begin{itemize}

    \item \textbf{Server:} The server is the coordinator of the protocol and takes following responsibilities, a) \textit{preparing} the entangled \textit{qubits}; b) \textit{distributing} these {qubits} to all participants; c) and \textit{decoding} the aggregated \textit{global model} based on the qubits sent back by the participants (see \textit{Quantum Aggregation Protocol} in Sect. \ref{Sect:QSA}).
    
    \item \textbf{Participant:}  Participants refer to all parties who participate in a Federated Learning session.
    Each participant has their own \textit{private data} and may use these data to train a \textit{local model} e.g. a logistic regression model or a neural network for certain machine learning tasks such as image classification. 
    
    Note that participants are motivated to collaboratively build a \textit{global model} that has improved performance (e.g. higher accuracy) than that of their respective local models. Since different participants may have different sizes of private data, in terms of different numbers of data samples. Therefore, the extent varies by which the global model performance improves relative to different local models (see Sect. \ref{Sect:Experiment} for experimental results).  
    For the sake of data privacy, on the other hand, participants are not willing (or not allowed) to disclose their private data to other participants or the server who join the same federated learning session.  To this end, participants are well motivated to take certain measures to protect private data from being espied by \textit{semi-honest} parties (see \textit{threat model} in Sect. \ref{subsect:threat} and security measure in Sect. \ref{Sect:Security}).

    \item \textbf{Quantum Channel:} \textit{quantum communication channels} are used to transmit \textit{qubits} encoding private model parameters between participants and the server.
    Due to the unique physical properties of Quantum communication channels,  semi-honest misbehaves who attempt to steal private model parameters by wiretapping quantum channels can be caught and prevented on spot (see Security Analysis in Sect. \ref{Sect:Security}).
    
    \item \textbf{Classical Channel:} \textit{classical communication channels} are used to transmit auxiliary signals that assist to establish quantum channels etc. In the centralized architecture illustrated in Sect. \ref{subsubsect:centralized}, the classical channel is used not only to transmit the auxiliary signals of the quantum channel, but also to dispatch the global model to participants.
    In a decentralized architecture in Sect. \ref{subsubsect:decentralized}, classical channels are only used to transmit quantum channel auxiliary signals (see Fig. \ref{QFL_structure}).
\end{itemize}

We illustrate below two different architectures of the proposed QFL system, namely a \textit{centralized} architecture and a \textit{decentralized} architecture. Briefly, the two architectures are similar in terms of the optimization of local models, but are different in the way with which the aggregated model are transmitted to different participants. Consequently, the \textit{security}
of the aggregated global model are fundamentally different in two architectures.

\subsubsection{A Centralized Architecture}\label{subsubsect:centralized}

Fig. \ref{QFL_structure} (a) illustrates the centralized architecture in which the server is only responsible for coordinating the entire protocol and does not participate in the local model training. 
Each participant use its local data to train the local model which is subsequently transmitted to the server through the \textit{quantum secure aggregation} (QSA). Note that parameters of the local model are decoded and aggregated completely in the quantum states. Finally, the server extracts parameters of the global model up to the desired precision by measuring received qubits states multiple times\footnote{Fig. \ref{var} shows that the number of measurements needs to be greater than $251$ to reduce variances of model parameters smaller than $10^-3$.} and broadcasts the aggregated global model to all participants through a classic channel.

In terms of security, the aggregation protocol in this centralized architecture can protect the private data as well as the local model of each participant (see Security Analysis in Sect. \ref{Sect:Security}).  Yet, the global model aggregated by the server broadcast through the classical channel still brings potential risks, with which a semi-honest server may launch \textit{model inversion} attack to infer from the aggregated model statistical information about private data of participants \cite{hitaj2017deep,melis2019exploiting,nasr2019comprehensive}.  This risk is mitigated in the decentralized architecture (see Sect. \ref{subsubsect:decentralized}). 

In terms of efficiency, the centralized architecture admits a fast global model convergence since the aggregated model is broadcasted to all participants after each iteration of the local model updating, allowing local models to be synchronized more frequently (see analysis in Sect. \ref{subsect:time-analysis}). 

\subsubsection{A Decentralized Architecture}\label{subsubsect:decentralized}

Fig. \ref{QFL_structure} (b) illustrates the decentralized architecture in which the server is no longer needed. Instead, the responsibility of the server is assigned to each participants in turn. That is to say, for each iteration of the federated learning, only one participant is designated as the server to coordinate the secure aggregation protocol (i.e. to prepare qubits, sends and receive qubits conveying information about the aggregated model). The designated participant is then able to extract the aggregated model from which the participant continues its local model learning. 

On the other hand, the rest of participants acts as participants which only send qubits convey information about the local model to the designated party.  Note that these participants are unable to update with the aggregated model. They merely continue optimizing the local model until their turn to be designated as the server. It turns out the updating with the aggregated model in multiple iterations still ensures the convergence of the aggregated model but at a slower convergence rate (see experiment results in Sect. \ref{Sect:Experiment})\footnote{It is worthy mentioning that the approach to update local models with the server in multiple iterations is also adopted in  federated learning with classical communication channels, for the sake of improved efficiency e.g. when Homomorphic Encryption (HE) is adopted \cite{BatchCrypt}}.

In terms of security, this decentralized architecture allows both the local model and the aggregated model always be encoded in the Quantum Channels, resulting in more secure protection of privacy information as compared with the centralized architecture (see security analysis in Sect. \ref{Sect:Security}).

\subsection{Local models} 

Like homomorphic encryption, the quantum aggregation protocol supports all machine learning models with parameters. In this article, we take multinomial logistic regression (LR)\cite{LR} and convolutional neural network (CNN)\cite{CNN} as examples to explore the performance of the quantum aggregation protocol for classical machine learning models. In addition, we also experimentally demonstrate that the quantum aggregation protocol can be used for quantum neural network (QNN).
We briefly illustrate these local models as follows and refer to Appendix \ref{Sect:QNN} for elaborated accounts of a QNN model.
\begin{itemize}
    \item The LR mode is a supervised learning algorithm commonly used for binary classification, and we adopt its variant for multi-class classification. We use $N$ independent binary logistic regressions to realize multinomial logistic regression, where $N$ is the number of data labels. The $i$-th binary logistic regression outs the probability of the input data is predicted as the $i$-th label \cite{mlr}.
    
    Implementation-wise, the above LR model is implemented as a two-layer neural network, with the number of neurons in the input layer being the dimension of a single data (i.e. 768), and the number of neurons in the output layer being the number of labels (i.e. 10). 
    
    \item The convolution neural network (CNN) model adopted in our experiments is an 18-layer Resnet model  \cite{resnet} which demonstrated superior classification accuracy performances e.g. on the Imagenet datasets.  It was used to classify each image in CIFAR10 datasets and output one of ten designated labels (i.e. airplane, automobile and bird etc. see \cite{cifar10}). 
    
    \item The QNN model is a deep quantum circuit network composed of blocks with parametric single qubit gates and two-qubits gates. 
    The input of QNN is the quantum state encoded by the amplitude of the data\cite{ampliencode}, and the output is the projection of the qubits on the z-axis. Due to the limitation of the computing power of current quantum computers, we conducted related experiments using simulators. Implementation details about QNN's are shown in Appendix \ref{Sect:QNN}.

\end{itemize}

\subsection{Quantum Aggregation Protocol}\label{Sect:QSA}

Quantum secure aggregation protocol can be used to protect the data security of various participants. We propose a quantum secure aggregation protocol based on GHZ-state, which can be used to calculate the sum of parameters provided by each participant without revealing the parameter information of any single participant. Here we assume that the aggregation server (which can be also served by one participant in turn) and the participants are both semi-honest.\\

\begin{itemize}
\item \textbf{Distribution:} the aggregation server generates a GHZ-state composed of $N$ qubits (Eq. \ref{GHZ_state}), and distributes these $N$ qubits to $N$ participants through quantum channels. \\

\begin{equation}
    \frac{1}{\sqrt{2}}(|000...0\rangle+|111...1\rangle)
    \label{GHZ_state}
\end{equation}
\item \textbf{Encoding:} When the $i$-th participant receives the qubit. After they confirm that the transmission is secure, he applies $R_Z(\theta_i)$ gate, where $\theta_i$ denotes the normalized parameter.\\

\item \textbf{Sending back:} Then each participant throws the encoded qubit back to the server. Till now, every participant encode his model parameter on the entangled state (Eq. \ref{encodeGHZ}).\\

\begin{equation}
    \frac{1}{\sqrt{2}}(|000...0\rangle+\exp(i(\Sigma_{n=1}^{N}\theta_n))\ket{111...1})
    \label{encodeGHZ}
\end{equation}
\item \textbf{Measurement:} After the server receives the encoded qubits, the server can decode the sum of $\theta_i$ by simply adapting $CNOT$-gates and $Hadamard$-gates (Eq. \ref{finalGHZ}) and then measures the first qubit, it will get 0 with a $\frac{1+cos(\Sigma_{n=1}^{N}\theta_n)}{2}$ probability. Therefore, by repeating the procedure, the server will get the estimation of $\Sigma_{n=1}^{N}\theta_n$.
\end{itemize}
\begin{equation}
\begin{aligned}
        H_1CNOT_{1,2}CNOT_{2,3}...CNOT_{N-1,N}\\
        (\frac{1}{\sqrt{2}}(\ket{000...0}+\exp(i(\Sigma_{n=1}^{N}\theta_n))\ket{111...1}))\\
        =\frac{1}{2}((1+\exp(i(\Sigma_{n=1}^{N}\theta_n))\ket{000...0}\\
        +(1-\exp(i(\Sigma_{n=1}^{N}\theta_n))\ket{000...1})
\end{aligned}
\label{finalGHZ}
\end{equation}

\begin{algorithm}
    \caption{QuantumSecureAggregation. The $N$ participants are named as $\left\{ C_i, i=1,2,...,N\right\}$ ; $C_i$'s private parameter is named as $\theta_i$; aggregation server is named as $S$, $n$ is the number of repetitions.}
    \label{qsa_algo}
    \begin{algorithmic}[1]
        \REPEAT
        \STATE Server $S$ entangles qubits $\left\{q_1, q_2, ... , q_N \right\}$ into state  $\frac{1}{\sqrt{2}}(|000...0\rangle+|111...1\rangle)$
        \REPEAT 
            \STATE $S$ sends $q_i$ to $C_i$ through \textbf{\underline{quantum channel}}
        \UNTIL All qubits are sent to participants
        \REPEAT
            \STATE $C_i$ applies $R_z(\theta_i)$ gate on $q_i$ (Encoding)
            \STATE $C_i$ sends $q_i$ to $S$ through \textbf{\underline{quantum channel}}
        \UNTIL All participants send back the qubits
        \STATE $S$ decodes all qubits and measure $q_1$, gets $0$ or $1$
        \UNTIL repeat $n$ times
        \STATE $S$ counts the number of occurrences of $q_1=0$ as $f_0$ and estimates $\Sigma_{i=1}^N \theta_i$ from 	$\arccos{(2f_0 - 1)}$ (Decoding)
        \STATE $S$ broadcasts the result to each participants through classical channel
    \end{algorithmic}
\end{algorithm}
The pseudocode of the protocol is shown in Algo. \ref{qsa_algo}.  This protocol can also transform to a decentralized version. In each round of aggregation, the protocol can choose $1$ participant to play the role of server. The chosen participant generates entangled qubits and sends them to other participants. Then other participants encode their information in the same way and send back to the chosen participant. Finally, the chosen participant decode the information without sharing with others. The decentralized protocol guaranteed that private information will never appear in the classic channel. The pseudocode of the protocol is shown in Algo. \ref{serverless_qsa_algo}.\\
\begin{algorithm}
    \caption{DecentralizedQuantumSecureAggregation. The $N$ participants are named as $\left\{ C_i, i=1,2,...,N\right\}$ ; $C_i$'s private parameter is named as $\theta_i$; $n$ is the number of repetitions; $k \leq n$ is the current round; $M_k$ is the chosen participant who plays the role of server.}
    \label{serverless_qsa_algo}
    \begin{algorithmic}[1]
        \REPEAT
        \STATE $M_k$ entangles qubits $\left\{q_1, q_2, ... , q_N \right\}$ into state  $\frac{1}{\sqrt{2}}(|000...0\rangle+|111...1\rangle)$
        \REPEAT 
            \STATE $M_k$ sends $q_i$ to $C_i$ through \textbf{\underline{quantum channel}}
        \UNTIL All qubits are sent to participants
        \REPEAT
            \STATE $C_i$ applies $R_z(\theta_i)$ gate on $q_i$ (Encoding)
            \STATE $C_i$ sends $q_i$ to $M_k$ through \textbf{\underline{quantum channel}}
        \UNTIL All participants send back the qubits
        \STATE $M_k$ decodes all qubits and measure $q_1$, gets $0$ or $1$
        \UNTIL repeat $n$ times
        \STATE $M_k$ counts the number of occurrences of $q_1=0$ as $f_0$ and estimates $\Sigma_{i=1}^N \theta_i$ from 	$\arccos{(2f_0 - 1)}$ (Decoding)
    \end{algorithmic}
\end{algorithm}

\subsubsection{Efficiency and time cost analysis}\label{subsect:time-analysis}

The proposed QSA scheme is not only secure but also efficient, in the sense that it does not require computationally demanding encryption methods e.g. homomorphic encryption to protect private model information.  Instead, model parameters are encoded by a number of entangled qubits which are sent from participants to the server for secure model aggregation (see Sect. \ref{Sect:QSA}).
\begin{align}\label{time_estimate}
    T \approx2(NMt_g+t_{net})
\end{align}

in which $N$ is the number of participants and $M$ is the number of entangled qubits to be sent by each participant to encode model parameters. $t_g (\sim 22 \mu s)$ is the time cost of operating a quantum gate and $t_{net} (\sim 1 ms)$ is the time cost of spreading or returning qubits through quantum channel \cite{quantum_channel}.

\begin{figure}
    \centering
    \includegraphics[scale=0.65]{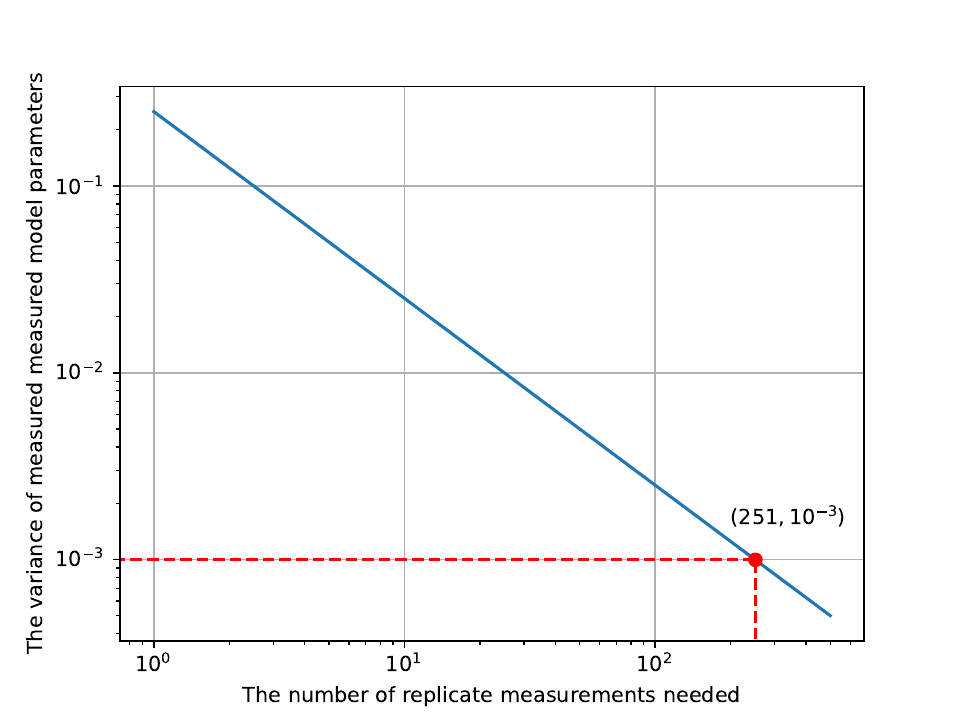}
    \caption{The relationship between the precision (variance) of measured model parameters and the number of replicate measurements needed  (X-axis: the number of replicate measurements needed; Y-axis: the variance of measured measured model parameters).
    }
    \label{var}
\end{figure}

Note that the number of entangled qubits $M$ is a crucial parameter that influences both the time cost and the precision of the measured model parameters.  On one hand, the precision increases with the number of replicate measurements $M$ and the variance of measured parameters decreases proportionally with the increasing of $M$ (see Fig. \ref{var}).  On the other hand, the time cost increases linearly with $M$.  A reasonable trade-off thus can be achieved by taking e.g. $M \approx 251$ which leads to a measurement error variance lower than $10^{-3}$ and the minor model performance degradation incurred by imprecision in model parameters is acceptable \cite{variance}\footnote{Here we only consider the variance of the results due to the randomness of the measurements and do not consider other noises in quantum computing and quantum networks. When $p=\frac{1}{N}\sum_{i=1}^N \theta_i = 0.5$, the variance of results $\sigma^2 = \sum_{i=0}^{M}C_{M}^{i}p^{i}(1-p)^{(n-i)}(\frac{i}{M}-p)^2$ is max. When $M$ is larger than $251$, $\sigma^2$ will be less than $10^{-3}$.}. 

\subsection{Security Analysis }\label{Sect:Security}

In this section, we analyse the security of the quantum secure aggregation protocol in both the centralized and decentralized architectures (see Sect. \ref{subsect:architect}). 

It is shown that the proposed protocol is secure against both external and participant attacks. 
Note that the server and the participants are assumed to be semi-honest, that is to say, they will follow the steps of the protocol, but may steal other participants’ private inputs based on the information they already have.

\subsubsection{Decoy states}
Decoy state technology can be used to detect whether the quantum channel has been eavesdropped. The principle is that when qubits of GHZ states are sent through a quantum channel, several decoy qubits, each randomly from $\{|0\rangle, |1\rangle, |+\rangle, |-\rangle\}$, are randomly inserted among them and sent. After the receiver receives these qubits, it measures the decoy qubits based on the sender's information about the positions and bases of decoy qubits (transmitted over a classical channel) and returns the result to the sender. The error rate calculated by the sender can determine whether the transmission has been eavesdropped.
To obtain information about participants' private inputs, an external malicious adversary has to perform some operations (including measurements) on the qubits that consist of the GHZ state. But such behaviors may affect the decoy states as the decoy states and the qubits from the GHZ-state are indistinguishable, leading to errors when the participants and the server compare the measurement results with the initial decoy states, which indicates that there exist  external malicious adversaries.

\subsubsection{Analysis of external attacks}

Let us first show that our protocol can resist external attacks in a centralized architecture. 
The decoy states $\{|0\rangle, |1\rangle, |+\rangle, |-\rangle\}$ can be utilized during the quantum communication between the server and the participants to detect the eavesdropping in the following way: 
\begin{enumerate}[label=\roman*.]
\item the server sends the qubit of GHZ-state with decoy states to the $i$-th participant; 

\item when the $i$-th participant receives the qubit of GHZ-state with decoy states, the server tells the $i$-th participant the positions of the decoy states and the basis of each decoy states; 

\item the $i$-th participant measures the corresponding decoy states with proper bases and shows the results to the server; 

\item the server computes the error rate based on these results to check whether the quantum transmission is secure. 
\end{enumerate}
When the encoded qubit of GHZ-state is sent back to the server, this method is also employ to ensure the security of the quantum transmission. This idea comes from the first quantum key distribution protocol by Bennett and Brassard \cite{QKD} that has been proven to be unconditional secure \cite{QKDproof}. Our protocol is thus secure against various external attacks, such as the intercept-resend attack, the  measurement-resend attack, and the entanglement-measurement attack. We take the measurement-resend attack as an example. Suppose there are $d$ decoy states used for eavesdropping detection between one participant and the server. For each decoy state affected by the eavesdropper Eve, Eve is able to escape the detection with probability $3/4$. $d$ decoy states get Eve caught with probability $1-(3/4)^{d}$ that will be approaching one when $d$ is large enough. 

Without loss of generality, the external eavesdropper can launch a general attack that can be described as
\begin{equation}\label{eq.zero}
\begin{array}{c}
    U_E|0\rangle|e\rangle  = |0\rangle|e_{00}\rangle+|1\rangle|e_{01}\rangle,\\
    U_E|1\rangle|e\rangle = |0\rangle|e_{10}\rangle+|1\rangle|e_{11}\rangle,
\end{array}
\end{equation}
where $U_E$ is a unitary operation and $|e_{ij}\rangle, i,j \in \{0,1\}$ are ancilla states. We will show that in order to pass the eavesdropping detection, the final state shared by the server and Eve is a product state of the GHZ state and an ancilla state.

As the decoy state are randomly chosen from $\{|0\rangle, |1\rangle, |+\rangle, |-\rangle\}$, its density operator is $\frac{I}{2}$ which is the same as the quantum state sent from the server to a participant. Namely, Eve cannot distinguish these two kinds of states.

On the one hand, if the decoy state is from $\{|0\rangle, |1\rangle\}$ and Eve escapes the detection, the following equation
\begin{equation}
    |e_{01}\rangle=|e_{10}\rangle=0,
\end{equation}
should hold, where $0$ is a zero vector, due to the equation (\ref{eq.zero}).

On the other hand, if the decoy state is from $\{|+\rangle, |-\rangle\}$, we have
\begin{equation}
\begin{array}{lll}
    U_E|+\rangle|e\rangle&=&U_E(\frac{1}{\sqrt{2}}(|0\rangle+|1\rangle)|e\rangle)  \\
     & =&\frac{1}{2}(|+\rangle (|e_{00}\rangle+|e_{01}\rangle+|e_{10}\rangle|+e_{11}\rangle)\\
     & &+|-\rangle (|e_{00}\rangle-|e_{01}\rangle+|e_{10}\rangle-|e_{11}\rangle)
\end{array}
\end{equation}
and
\begin{equation}
\begin{array}{lll}
    U_E|-\rangle|e\rangle&=&U_E(\frac{1}{\sqrt{2}}(|0\rangle-|1\rangle)|e\rangle)  \\
     & =&\frac{1}{2}(|+\rangle (|e_{00}\rangle+|e_{01}\rangle-|e_{10}\rangle-|e_{11}\rangle)\\
     & &+|-\rangle (|e_{00}\rangle-|e_{01}\rangle-|e_{10}\rangle+|e_{11}\rangle).
\end{array}
\end{equation}

In order to pass the eavesdropping detection, we obtain
\begin{equation}\label{eq.plus}
\begin{array}{c}
     |e_{00}\rangle-|e_{01}\rangle+ |e_{10}\rangle -|e_{11}\rangle=0,  \\
     |e_{00}\rangle+|e_{01}\rangle- |e_{10}\rangle -|e_{11}\rangle=0, 
\end{array}
\end{equation}
where $0$ is a zero vector.

Equations (\ref{eq.zero}) and (\ref{eq.plus}) imply
\begin{equation}\label{eq.plus.0}
\begin{array}{l}
     |e_{00}\rangle=|e_{11}\rangle,  \\
     |e_{01}\rangle=|e_{10}\rangle=0. 
\end{array}
\end{equation}

That is to say, the ancilla state cannot be entangled with the quantum state sent from the server to the participant. Therefore, Eve obtains nothing about participants' private inputs.

Note that Trojan horse attacks, such as the delay-photon Trojan horse attack and the invisible photon eavesdropping Trojan horse attack, may exist in our protocol. However, the photon number splitter and the optical wavelength filter devices can be used to detect such attacks.\cite{QSSsecurity1, QSSsecurity2}

\subsubsection{Analysis of participant attacks}.

Let us now move on to the analysis of the participant attacks. We assume that the server and the participants are semi-honest. They should loyally follow the procedure of the protocol and they cannot collude with each other.

For the attacks from the server, the server may first send a quantum state $|+\rangle$ to a participant, say $P_i$,  and then $P_i$ encodes their private input $\theta_i$ to the $|+\rangle$, getting the final state $|\varphi\rangle=RZ(\theta_i)|+\rangle$. When $P_i$ sends $|\varphi\rangle$ to the server, the server can easily obtain $\theta_i$. However, this task cannot be finished as the server is assumed to be semi-honest, and it cannot send fake quantum states.

\textbf{No motivation for participants to attack}.  For the attacks from dishonest participants, they may try to catch the encoded quantum state sent from honest participants to the server to obtain information about their private inputs $\theta$. But they will fail because the encoded quantum state is $\frac{1}{2}I$ that contains nothing about $\theta$. 
Let us give the details. As we can see from the proposed protocol, the original GHZ-state prepared by the server is $|\varphi\rangle=\frac{1}{\sqrt{2}}(|000\ldots 0\rangle_{123\ldots N} + |111\ldots 1
\rangle_{123\ldots N})$. Each qubit from this state becomes $\rho_i=tr_{12(i-1)(i+1)N}(|\varphi\rangle\langle\varphi|)=\frac{I}{2}$ $(i=1, 2, \ldots,N)$. Similarly,
the GHZ-state encoded by all participants through $\{\theta_{i}|i=1, 2, \ldots, N\}$ reads $|\varphi^{\prime}\rangle=\frac{1}{\sqrt{2}}(|000\ldots 0\rangle_{123\ldots N} +(\sum_{i=1}^{N}\theta_{i} )|1\ldots 1\rangle_{123\ldots N})$. Each qubit from that encoded state turns into $\rho_i^{\prime}=tr_{12(i-1)(i+1)N}(|\varphi^{\prime}\rangle\langle\varphi^{\prime}|)=\frac{I}{2}$ $(i=1, 2, \ldots,N)$.
That is to say, each qubit from the encoded GHZ-state reveals nothing about each participant's private input and the dishonest participants who 
cannot launch collusion attacks will fail to steal it.

\subsubsection{Analysis of security in a decentralized architecture}
As seen in the previous analysis, the security of the aggregation protocol in a centralized architecture is guaranteed by the detecting method of decoy states. Decentralized architecture can also be reduced to the scenario that one of the participants plays the role of the main sever. Therefore, the security in a decentralized architecture remains the same as that in a centralized one.

\subsubsection{Security analysis against server side attacks}
A malicious server may distribute fake quantum states instead of GHZ states to participants to steal their private inputs. However, this attack will be detected if all participants cooperate with each other. Specifically, a GHZ state can be rewritten as $\frac{1}{\sqrt{2}} (\ket{000...0} + \ket{111...1})$ or $\frac{1}{\sqrt{2}^{n-1}}\sum \ket{a_1a_2...a_n}$, where $a_i\in {\ket{-}, \ket{+}}$ and the number of $\ket{-}$ is even. If all participants measure their quits from GHZ in the computational basis, they should get the same results. If all participants measure their qubits from GHZ in the diagonal basis, the numbers of $\ket{-}$ should be even. This fact provides a way to detect whether a genuine GHZ state is distributed among participants.

\section{Experiment results}\label{Sect:Experiment}

This section illustrates experimental results of applying the proposed QFL framework to a typical machine learning task i.e. image classification using different image datasets.  
Experimental settings are briefly summarized as follows. 

    The \textit{number of FL participants} are set as 3, 5 or 10 in different experiments. 
    The \textit{types of local models} include the well-known logistic regression (LR) \cite{mlr}, convolution neural network (CNN) \cite{resnet} as well as Quantum Neural Network (QNN) (see Appendix \ref{Sect:QNN}).  Note that the LR and CNN are typical \textit{classical} machine learning models which are widely used for a variety of applications. 
    The proposed quantum secure aggregation is applied to different typed of local models. 
    Two different \textit{architectures} i.e. \textit{centralized} and \textit{decentralized} federated learning are adopted to evaluate performances of aggregated global models for image classification tasks.  Both MNIST \cite{mnist} and CIFAR10 \cite{cifar10} image datasets are used for experiments illustrated below. 

\subsection{Classical machine learning models}

We first experimentally evaluate performances of the aggregated models when participants' local model are classical machine learning models i.e. LR and CNN.  Note that different sizes of the training datasets (in terms of the number of data samples) are assigned to different participants, and this non-IID dataset setting  well simulates situations in real life where different organizations, e.g. hospitals or banks, may have drastically different sizes of datasets. 

\begin{figure*}
    \centering
    \includegraphics[width=\linewidth]{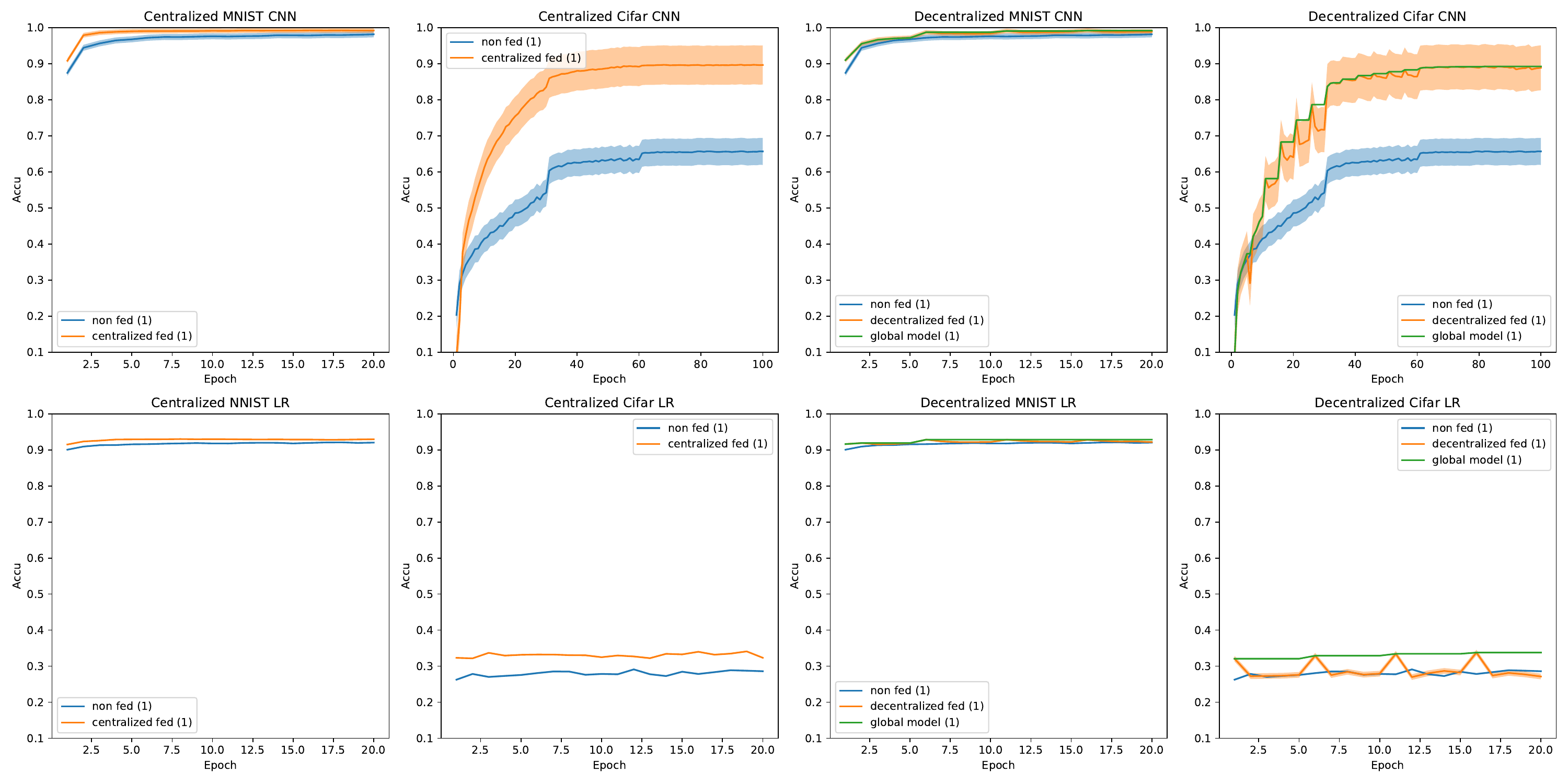}
    \caption{The figure shows the MNIST dataset and the CIFAR10 dataset, as well as the experimental performance of the LR model and the CNN model in the centralized and decentralized architectures and the corresponding $95\%$ confidence intervals. The blue line in the figure represents the effect of the model trained with only local data on the local test set when participant 1 with the smallest data set ($10\%$) does not participate in federated learning. The orange line represents the performance of the global model on its local test set after participant 1 joins federated learning. It is important to point out that in decentralized federated learning, because the non-IID local data can lead to concept drift of the received model, we choose the best performing model (green line) so far as the global model.}
    \label{classical_model_result}
\end{figure*}

Fig. \ref{classical_model_result} illustrates model accuracies for image classification evaluated with MNIST and CIFAR10 datasets. First, it was shown that  the global model performances are improved, to various extents, as compared to that of participants' local models. In particular, model accuracies increase significantly (e.g. from 65\% to almost 90\%)  for those participants whose local datasets are of limited sizes. It is this type of model performance improvements that motivate participants to participate a federated learning session. 
Second, it was shown that the number of replicate measurements is 251. This measurement step took up to $0.33$ seconds according to eq. \ref{time_estimate}. Therefore, it is shown that the proposed Quantum Secure Aggregation protocol is \textit{efficient} in terms of the number of replicate measurements and the time cost needed in measuring global model parameters up to required precision. 
Third, the QSA protocol also addresses the concern of privacy leakages of training data as demonstrated in Sect. \ref{Sect:Security}. The superior performances of the proposed QSA protocol  justifies it as a practical solution for federated learning.

\subsection{QNN}
To explore the performance of QSA and QNN in FL under real scenarios, we conduct numerical experiments simulating real-world situations. We explore the effect of federated learning using QSA and QNN on participants' local models by changing the way the data is divided.
\begin{figure}
    \centering
    \includegraphics[width=\linewidth]{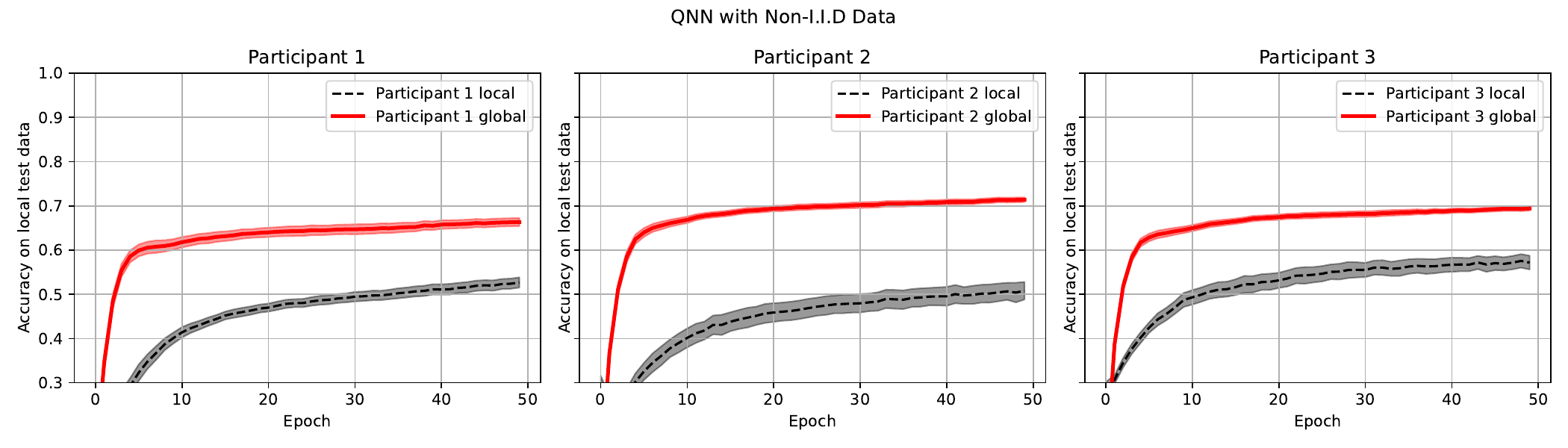}
    \caption{We divide the MNIST dataset into Non-I.I.D datasets with a scale of $1:2:3$, which are used as private data for participants 1, 2, and 3, respectively. Each participant participates in a model aggregation after training a round of QNN locally. It can be found that the global model far outperforms the local model that does not participate in federated learning on the private datasets of each participant. The experimental results show that QSA is also applicable to the QNN model.}
    \label{amazingresult}
\end{figure}

In Fig. \ref{amazingresult}, we show the improvement of QNN in federated learning under unequal data size and different distribution of labels. We performed 30 experiments for each condition and plotted curves with confidence intervals. We found that after federated learning, the global model performed better on each of the three parties' respective test sets than the models trained by the three parties individually. In particular, it is most helpful to the participants with a small amount of data, and less helpful to the participants with the largest amount of data. It shows that federated learning is beneficial to the party with less data.
In addition, in the case of different distributions of labels, some participants may have inherent defects in the data, resulting in poor performance of the trained model. However, after federated learning, each participant contributes their own unique data, making the global model of federated learning close to the upper limit of accuracy that QNN can achieve.\\

\section{Discussion}\label{sect:disc}

We discuss below a number of open issues concerning the comparison of the proposed QSA with classical secure aggregation methods, the feasibility of practical implementations of the proposed QSA scheme.

\textbf{Comparison with the classical secure aggregation}: security analysis in Sect. \ref{Sect:Security} shown that it is impossible to eavesdrop information transmitted via quantum channels without being noticed and stopped immediately.  As compared with two representative classical secure aggregation methods i.e. differential privacy (DP) \cite{dwork2006differential, DP1, DP2, DP3, DP4} and homomorphic encryption \cite{FHE, HEsurvey, PHE, FHE}, the proposed QSA scheme is advantageous in that the amount of information might be leaked to semi-honest adversaries is virtually zero i.e. achieving \textit{perfect secrecy}, and there is no need to maintain private keys and distribute public keys as in HE. In terms of efficiency and complexity, the required number of repeated measurements of qubits in QSA be around $251$ to maintain an acceptable level of model accuracy and the total complexity increases \textit{linearly} with this crucial parameter (see Sect. \ref{subsect:time-analysis}). While for classical encryption, it often requires the key lengths be over a thousand bits and the computational complexity involved increases exponentially with the key length. To this end, we view the QSA scheme a promising alternative to existing secure aggregation approaches provided that enabling techniques of \textit{quantum internet} become feasible (see Discussion below). 

\textbf{Practical implementations of the QSA scheme}: one crucial issue concerning the applicability of the proposed Quantum state-based Secure Aggregation (QSA) is rooted in the feasibility of techniques enabling \textbf{teleporting entangled states} and \textbf{quantum-gate over long distances}, which are adopted, respectively, in the \textit{distribution} and \textit{encoding} steps of the QSA scheme (see Sect. \ref{Sect:QSA}).  
Such two techniques are also core techniques to realize the long-term vision of \textit{quantum internet}, which aims to revolutionize the classical \textit{internet} with the superior computational power of quantum computer and the high security of information transmitted via quantum communication channels \cite{quantum_channel, QInternet1, QInternet2}.  
Implementation-wise, it was shown that teleporting entangled states and quantum-gate can be achieved over a distance over 60 meters in everyday environments on ground\cite{quantum_channel}. 
Moreover, it is worth mentioning that Pan \textit{et. al.} \cite{QInternet1, QInternet2} has empirically demonstrated that teleporting of entangled states can be achieved over a distance of more than a thousand of kilometers between a ground station and a satellite in space. Bearing in mind the rapid progresses of these required techniques, we are optimistic about the practical implementation of the proposed QSA scheme in a real-life federated learning scenario in near future.

\textbf{Byzantine robust}: in the face of malicious adversaries i.e. Byzantine attackers, who aim to damage global model performances by submitting poor-performed local models, the protection provided by Quantum Secure Aggregation (QSA) is effective against those attacks that are launched on the quantum channels. The security of QSA lies in the fact that such interference misbehaves can be detected by a number of honest parties and immediately stopped (see security analysis in Sect. \ref{Sect:Security}). On the other hand, however, in case that a local model is compromised by adversaries in the classical computing regime, QSA is unable to notice such abnormality and one has to resort to classical Byzantine Resilient methods such as Krum \cite{blanchard2017machine} and trimmed mean \cite{yin2018byzantine}.

\section{Conclusion}\label{sect:conclu}

In this article, we propose a Quantum-entangled state-based Secure Aggregation method (QSA), analyze its security, and build a new federated learning framework (QFL) based on it. Such a QFL framework has demonstrated significant performance boosting capability to different types of machine learning models. Performance improvements are especially pronounced when sizes of local datasets are limited and the Quantum Neural Network is employed. 
The QFL framework is also secure in protecting private model parameters as well as private data from being disclosed to semi-honest adversaries. Moreover, any misbehaves to eavesdrop private information can be caught immediately and stopped. To this end, we view the proposed QSA protocol an effective and promising security mechanism for federated learning in which the protection of data and model is of paramount importance, and we hope that the present article, in tandem with follow up works, will make impactful contributions to federated learning research.

\bibliographystyle{unsrt}
\bibliography{bibliography}

\begin{thebibliography}{10}

\bibitem{PHE}
Pascal Paillier.
\newblock Public-key cryptosystems based on composite degree residuosity classes.
\newblock In {\em International conference on the theory and applications of cryptographic techniques}, pages 223--238. Springer, 1999.

\bibitem{FHE}
Craig Gentry.
\newblock Fully homomorphic encryption using ideal lattices.
\newblock In {\em Proceedings of the forty-first annual ACM symposium on Theory of computing}, pages 169--178, 2009.

\bibitem{DP1}
Cynthia Dwork.
\newblock Differential privacy: A survey of results.
\newblock In {\em International conference on theory and applications of models of computation}, pages 1--19. Springer, 2008.

\bibitem{DP2}
Kang Wei, Jun Li, Ming Ding, Chuan Ma, Howard~H Yang, Farhad Farokhi, Shi Jin, Tony~QS Quek, and H~Vincent Poor.
\newblock Federated learning with differential privacy: Algorithms and performance analysis.
\newblock {\em IEEE Transactions on Information Forensics and Security}, 15:3454--3469, 2020.

\bibitem{DP3}
Aleksei Triastcyn and Boi Faltings.
\newblock Federated learning with bayesian differential privacy.
\newblock In {\em 2019 IEEE International Conference on Big Data (Big Data)}, pages 2587--2596. IEEE, 2019.

\bibitem{DP4}
Yoshinori Aono, Takuya Hayashi, Lihua Wang, Shiho Moriai, et~al.
\newblock Privacy-preserving deep learning via additively homomorphic encryption.
\newblock {\em IEEE Transactions on Information Forensics and Security}, 13(5):1333--1345, 2017.

\bibitem{HEsurvey}
Abbas Acar, Hidayet Aksu, A~Selcuk Uluagac, and Mauro Conti.
\newblock A survey on homomorphic encryption schemes: Theory and implementation.
\newblock {\em ACM Computing Surveys (Csur)}, 51(4):1--35, 2018.

\bibitem{BatchCrypt}
Chengliang Zhang, Suyi Li, Junzhe Xia, Wei Wang, Feng Yan, and Yang Liu.
\newblock $\{$BatchCrypt$\}$: Efficient homomorphic encryption for $\{$Cross-Silo$\}$ federated learning.
\newblock In {\em 2020 USENIX Annual Technical Conference (USENIX ATC 20)}, pages 493--506, 2020.

\bibitem{zhu_deep_2020}
Ligeng Zhu and Song Han.
\newblock Deep {Leakage} from {Gradients}.
\newblock In Qiang Yang, Lixin Fan, and Han Yu, editors, {\em Federated {Learning}: {Privacy} and {Incentive}}, pages 17--31. Springer International Publishing, Cham, 2020.

\bibitem{NFL}
Xiaojin Zhang, Hanlin Gu, Lixin Fan, Kai Chen, and Qiang Yang.
\newblock No free lunch theorem for security and utility in federated learning.
\newblock {\em arXiv preprint arXiv:2203.05816}, 2022.

\bibitem{QSS}
Mark Hillery, Vladim{\'\i}r Bu{\v{z}}ek, and Andr{\'e} Berthiaume.
\newblock Quantum secret sharing.
\newblock {\em Physical Review A}, 59(3):1829, 1999.

\bibitem{QML-2013}
Seth Lloyd, Masoud Mohseni, and Patrick Rebentrost.
\newblock Quantum algorithms for supervised and unsupervised machine learning, 2013.

\bibitem{QSVM-2014}
Patrick Rebentrost, Masoud Mohseni, and Seth Lloyd.
\newblock Quantum support vector machine for big data classification.
\newblock {\em Phys. Rev. Lett.}, 113:130503, Sep 2014.

\bibitem{QML_survey_Nat17}
Jacob Biamonte, Peter Wittek, Nicola Pancotti, Patrick Rebentrost, Nathan Wiebe, and Seth Lloyd.
\newblock Quantum machine learning.
\newblock {\em Nature}, 549(7671):195--202, sep 2017.

\bibitem{QNNspeedup_1}
Esma A{\"\i}meur, Gilles Brassard, and S{\'e}bastien Gambs.
\newblock Quantum speed-up for unsupervised learning.
\newblock {\em Machine Learning}, 90(2):261--287, 2013.

\bibitem{QNNspeedup_2}
Seth Lloyd, Masoud Mohseni, and Patrick Rebentrost.
\newblock Quantum algorithms for supervised and unsupervised machine learning.
\newblock {\em arXiv preprint arXiv:1307.0411}, 2013.

\bibitem{QNNspeedup_3}
Kerstin Beer, Dmytro Bondarenko, Terry Farrelly, Tobias~J Osborne, Robert Salzmann, Daniel Scheiermann, and Ramona Wolf.
\newblock Training deep quantum neural networks.
\newblock {\em Nature communications}, 11(1):1--6, 2020.

\bibitem{QFL_related_work_1}
Rui Huang, Xiaoqing Tan, and Qingshan Xu.
\newblock Quantum federated learning with decentralized data.
\newblock {\em IEEE Journal of Selected Topics in Quantum Electronics}, 28(4):1--10, 2022.

\bibitem{QFL_related_work_2}
Samuel Yen-Chi Chen and Shinjae Yoo.
\newblock Federated quantum machine learning.
\newblock {\em Entropy}, 23(4):460, 2021.

\bibitem{QFL_related_work_3}
Qi~Xia and Qun Li.
\newblock Quantumfed: a federated learning framework for collaborative quantum training.
\newblock {\em arXiv preprint arXiv:2106.09109}, 2021.

\bibitem{QFL_related_work_4}
Mahdi Chehimi and Walid Saad.
\newblock Quantum federated learning with quantum data.
\newblock In {\em ICASSP 2022 - 2022 IEEE International Conference on Acoustics, Speech and Signal Processing (ICASSP)}, pages 8617--8621, 2022.

\bibitem{QFL_related_work_5}
Weikang Li, Sirui Lu, and Dong-Ling Deng.
\newblock Quantum federated learning through blind quantum computing.
\newblock {\em Science China Physics, Mechanics \& Astronomy}, 64(10):1--8, 2021.

\bibitem{du2021learnability}
Yuxuan Du, Min-Hsiu Hsieh, Tongliang Liu, Shan You, and Dacheng Tao.
\newblock Learnability of quantum neural networks.
\newblock {\em PRX Quantum}, 2(4):040337, 2021.

\bibitem{du2022efficient}
Yuxuan Du, Zhuozhuo Tu, Xiao Yuan, and Dacheng Tao.
\newblock Efficient measure for the expressivity of variational quantum algorithms.
\newblock {\em Physical Review Letters}, 128(8):080506, 2022.

\bibitem{du2022quantum}
Yuxuan Du, Tao Huang, Shan You, Min-Hsiu Hsieh, and Dacheng Tao.
\newblock Quantum circuit architecture search for variational quantum algorithms.
\newblock {\em npj Quantum Information}, 8(1):1--8, 2022.

\bibitem{sheng2017distributed}
Yu-Bo Sheng and Lan Zhou.
\newblock Distributed secure quantum machine learning.
\newblock {\em Science Bulletin}, 62(14):1025--1029, 2017.

\bibitem{mcmahan2016federated}
H~Brendan McMahan, Eider Moore, Daniel Ramage, and Blaise~Ag{\"u}era y~Arcas.
\newblock Federated learning of deep networks using model averaging.
\newblock {\em arXiv preprint arXiv:1602.05629}, 2016.

\bibitem{mcmahan2017communication}
Brendan McMahan, Eider Moore, Daniel Ramage, Seth Hampson, and Blaise~Aguera y~Arcas.
\newblock Communication-efficient learning of deep networks from decentralized data.
\newblock In {\em Artificial Intelligence and Statistics}, pages 1273--1282. PMLR, 2017.

\bibitem{konevcny2016federated}
Jakub Kone{\v{c}}n{\`y}, H~Brendan McMahan, Daniel Ramage, and Peter Richt{\'a}rik.
\newblock Federated optimization: Distributed machine learning for on-device intelligence.
\newblock {\em arXiv preprint arXiv:1610.02527}, 2016.

\bibitem{konevcny2016federated_new}
Jakub Kone{\v{c}}n{\`y}, H~Brendan McMahan, Felix~X Yu, Peter Richt{\'a}rik, Ananda~Theertha Suresh, and Dave Bacon.
\newblock Federated learning: Strategies for improving communication efficiency.
\newblock {\em arXiv preprint arXiv:1610.05492}, 2016.

\bibitem{yang2019federated}
Qiang Yang, Yang Liu, Tianjian Chen, and Yongxin Tong.
\newblock Federated machine learning: Concept and applications.
\newblock {\em ACM Transactions on Intelligent Systems and Technology (TIST)}, 10(2):1--19, 2019.

\bibitem{DBLP:journals/ftml/KairouzMABBBBCC21}
Peter Kairouz, H.~Brendan McMahan, Brendan Avent, Aur{\'{e}}lien Bellet, Mehdi Bennis, Arjun~Nitin Bhagoji, Kallista~A. Bonawitz, Zachary Charles, Graham Cormode, Rachel Cummings, Rafael G.~L. D'Oliveira, Hubert Eichner, Salim~El Rouayheb, David Evans, Josh Gardner, Zachary Garrett, Adri{\`{a}} Gasc{\'{o}}n, Badih Ghazi, Phillip~B. Gibbons, Marco Gruteser, Za{\"{\i}}d Harchaoui, Chaoyang He, Lie He, Zhouyuan Huo, Ben Hutchinson, Justin Hsu, Martin Jaggi, Tara Javidi, Gauri Joshi, Mikhail Khodak, Jakub Kone{\v{c}}n{\'y}, Aleksandra Korolova, Farinaz Koushanfar, Sanmi Koyejo, Tancr{\`{e}}de Lepoint, Yang Liu, Prateek Mittal, Mehryar Mohri, Richard Nock, Ayfer {\"{O}}zg{\"{u}}r, Rasmus Pagh, Hang Qi, Daniel Ramage, Ramesh Raskar, Mariana Raykova, Dawn Song, Weikang Song, Sebastian~U. Stich, Ziteng Sun, Ananda~Theertha Suresh, Florian Tram{\`{e}}r, Praneeth Vepakomma, Jianyu Wang, Li~Xiong, Zheng Xu, Qiang Yang, Felix~X. Yu, Han Yu, and Sen Zhao.
\newblock Advances and open problems in federated learning.
\newblock {\em Found. Trends Mach. Learn.}, 14(1-2):1--210, 2021.

\bibitem{dwork2006differential}
Cynthia Dwork.
\newblock Differential privacy.
\newblock In {\em International Colloquium on Automata, Languages, and Programming}, pages 1--12. Springer, 2006.

\bibitem{dwork2014algorithmic}
Cynthia Dwork, Aaron Roth, et~al.
\newblock The algorithmic foundations of differential privacy.
\newblock {\em Foundations and Trends in Theoretical Computer Science}, 9(3-4):211--407, 2014.

\bibitem{abadi2016deep}
Martin Abadi, Andy Chu, Ian Goodfellow, H~Brendan McMahan, Ilya Mironov, Kunal Talwar, and Li~Zhang.
\newblock Deep learning with differential privacy.
\newblock In {\em Proceedings of the 2016 ACM SIGSAC conference on computer and communications security}, pages 308--318, New York, NY, USA, 2016. ACM.

\bibitem{GHZstate}
Daniel~M. Greenberger, Michael~A. Horne, and Anton Zeilinger.
\newblock Going beyond bell's theorem.
\newblock 2007.

\bibitem{hitaj2017deep}
Briland Hitaj, Giuseppe Ateniese, and Fernando Perez-Cruz.
\newblock Deep models under the gan: information leakage from collaborative deep learning.
\newblock In {\em Proceedings of the 2017 ACM SIGSAC conference on computer and communications security}, pages 603--618, 2017.

\bibitem{melis2019exploiting}
Luca Melis, Congzheng Song, Emiliano De~Cristofaro, and Vitaly Shmatikov.
\newblock Exploiting unintended feature leakage in collaborative learning.
\newblock In {\em 2019 IEEE symposium on security and privacy (SP)}, pages 691--706. IEEE, 2019.

\bibitem{nasr2019comprehensive}
Milad Nasr, Reza Shokri, and Amir Houmansadr.
\newblock Comprehensive privacy analysis of deep learning: Passive and active white-box inference attacks against centralized and federated learning.
\newblock In {\em 2019 IEEE symposium on security and privacy (SP)}, pages 739--753. IEEE, 2019.

\bibitem{LR}
David~R Cox.
\newblock The regression analysis of binary sequences.
\newblock {\em Journal of the Royal Statistical Society: Series B (Methodological)}, 20(2):215--232, 1958.

\bibitem{CNN}
Keiron O'Shea and Ryan Nash.
\newblock An introduction to convolutional neural networks.
\newblock {\em arXiv preprint arXiv:1511.08458}, 2015.

\bibitem{mlr}
Chanyeong Kwak and Alan Clayton-Matthews.
\newblock Multinomial logistic regression.
\newblock {\em Nursing research}, 51(6):404--410, 2002.

\bibitem{resnet}
Kaiming He, Xiangyu Zhang, Shaoqing Ren, and Jian Sun.
\newblock Deep residual learning for image recognition.
\newblock In {\em Proceedings of the IEEE conference on computer vision and pattern recognition}, pages 770--778, 2016.

\bibitem{cifar10}
Alex Krizhevsky, Geoffrey Hinton, et~al.
\newblock Learning multiple layers of features from tiny images.
\newblock 2009.

\bibitem{ampliencode}
Maria Schuld and Francesco Petruccione.
\newblock {\em Supervised learning with quantum computers}, volume~17.
\newblock Springer, 2018.

\bibitem{quantum_channel}
Severin Daiss, Stefan Langenfeld, Stephan Welte, Emanuele Distante, Philip Thomas, Lukas Hartung, Olivier Morin, and Gerhard Rempe.
\newblock A quantum-logic gate between distant quantum-network modules.
\newblock {\em Science}, 371(6529):614--617, 2021.

\bibitem{variance}
Ligeng Zhu, Zhijian Liu, and Song Han.
\newblock Deep leakage from gradients.
\newblock {\em Advances in Neural Information Processing Systems}, 32, 2019.

\bibitem{QKD}
Charles~H Bennett and Gilles Brassard.
\newblock Quantum cryptography: Public key distribution and coin tossing.
\newblock {\em arXiv preprint arXiv:2003.06557}, 2020.

\bibitem{QKDproof}
Peter~W Shor and John Preskill.
\newblock Simple proof of security of the bb84 quantum key distribution protocol.
\newblock {\em Physical review letters}, 85(2):441, 2000.

\bibitem{QSSsecurity1}
Fu-Guo Deng, Xi-Han Li, Hong-Yu Zhou, and Zhan-jun Zhang.
\newblock Improving the security of multiparty quantum secret sharing against trojan horse attack.
\newblock {\em Physical Review A}, 72(4):044302, 2005.

\bibitem{QSSsecurity2}
Xi-Han Li, Fu-Guo Deng, and Hong-Yu Zhou.
\newblock Improving the security of secure direct communication based on the secret transmitting order of particles.
\newblock {\em Physical Review A}, 74(5):054302, 2006.

\bibitem{mnist}
Yann LeCun and Corinna Cortes.
\newblock {MNIST} handwritten digit database.
\newblock 2010.

\bibitem{QInternet1}
H~Jeff Kimble.
\newblock The quantum internet.
\newblock {\em Nature}, 453(7198):1023--1030, 2008.

\bibitem{QInternet2}
Peter~C Humphreys, Norbert Kalb, Jaco~PJ Morits, Raymond~N Schouten, Raymond~FL Vermeulen, Daniel~J Twitchen, Matthew Markham, and Ronald Hanson.
\newblock Deterministic delivery of remote entanglement on a quantum network.
\newblock {\em Nature}, 558(7709):268--273, 2018.

\bibitem{blanchard2017machine}
Peva Blanchard, El~Mahdi El~Mhamdi, Rachid Guerraoui, and Julien Stainer.
\newblock Machine learning with adversaries: Byzantine tolerant gradient descent.
\newblock {\em Advances in Neural Information Processing Systems}, 30, 2017.

\bibitem{yin2018byzantine}
Dong Yin, Yudong Chen, Ramchandran Kannan, and Peter Bartlett.
\newblock Byzantine-robust distributed learning: Towards optimal statistical rates.
\newblock In {\em International Conference on Machine Learning}, pages 5650--5659. PMLR, 2018.

\bibitem{schuld2020circuit}
Maria Schuld, Alex Bocharov, Krysta~M Svore, and Nathan Wiebe.
\newblock Circuit-centric quantum classifiers.
\newblock {\em Physical Review A}, 101(3):032308, 2020.

\bibitem{overpara}
Martin Larocca, Nathan Ju, Diego Garc{\'\i}a-Mart{\'\i}n, Patrick~J Coles, and Marco Cerezo.
\newblock Theory of overparametrization in quantum neural networks.
\newblock {\em arXiv preprint arXiv:2109.11676}, 2021.

\bibitem{kingma2014adam}
Diederik~P Kingma and Jimmy Ba.
\newblock Adam: A method for stochastic optimization.
\newblock {\em arXiv preprint arXiv:1412.6980}, 2014.

\bibitem{QuantumTeleportation}
Charles~H Bennett, Gilles Brassard, Claude Cr{\'e}peau, Richard Jozsa, Asher Peres, and William~K Wootters.
\newblock Teleporting an unknown quantum state via dual classical and einstein-podolsky-rosen channels.
\newblock {\em Physical review letters}, 70(13):1895, 1993.

\end{thebibliography}

\newpage
\begin{appendices}
\section{Quantum Neural Network (QNN)}\label{Sect:QNN}

\begin{figure}
    \centering
    \includegraphics[width=\linewidth]{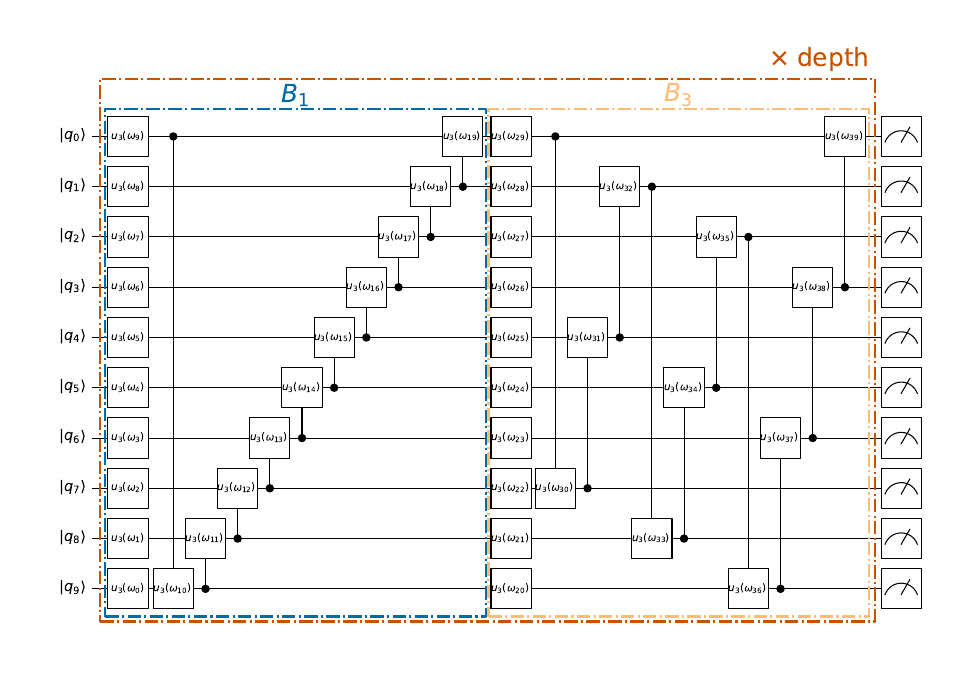}
    \caption{Structure of quantum neural network}
    \label{fig:qnn-structure}
\end{figure}

The basic concepts of quantum computation have been introduced in Sect. \ref{sect:preliminary}. This section will elaborate on the setup of quantum neural network (QNN). The structure of QNN is shown in Fig. \ref{fig:qnn-structure} which is composed of data embedding, parameterized circuits and measurements. Before discussing these three parts, it's helpful to summarize the connections between QNNs and artificial neural networks (ANNs):
\begin{enumerate}
    \item Data $x$: The image to be fed into ANNs is represented as a 3-dimensional array of shape $(\mathrm{Height},\mathrm{Width},\mathrm{Channel})$. For example the MNIST image is of the shape $(28,28,1)$. In QNNs, the image is encoded in the state vector $|q_0\rangle\otimes|q_1\rangle\otimes\cdots\otimes|q_9\rangle$ as shown in the leftmost of the Fig. \ref{fig:qnn-structure}. The detailed encoding strategy will be explained later.
    \item Parameters $\omega$: Most trainable parameters are wrapped in the fully-connected layers and convolutional layers in ANNs. Their counterpart in QNNs are the parameters $\omega$ in the parameterized quantum gates, such as $u_3(\omega)$ and $cu_3(\omega)$ gates shown in the Fig. \ref{fig:qnn-structure}.
    \item Probability prediction: In ANNs, the outputs after the softmax layer are treated as the probabilities of each label. Similarly, the measurement operations, the rightmost part in the Fig. \ref{fig:qnn-structure}, give the probability distribution over the whole label set and the highest one predicts the label of that input image.
\end{enumerate}

Various encoding strategies exist to map classical data into the quantum system. This work will focus on the amplitude encoding strategy for its efficiency in encoding high dimensional features and simplicity in implementing on the quantum simulator. Since the quantum state vector's length is required to be of the power of two for a $n$-qubit system in the computational basis, we need ``pad'' appropriate number of zeros to the end of the classical data. For MNIST data of $28\times 28=784$ dimensional vector and $10$-qubit system, extra $240$ $0$s are appended to the original data. Additionally, to satisfy the unit norm requirement of quantum state $|\langle \psi | \psi \rangle|^2=1$, an extra factor is used to scale the input data. The whole preprocessing to make the initial state vector is formulated as following:
\[ |q_0q_1\cdots q_9\rangle =\frac{1}{\sqrt{\chi}}\left[ p_1,p_2,p_3,\cdots ,p_{784},\underset{240}{\underbrace{0,\cdots ,0}} \right] \]
\[ \chi =\sum_{i=1}^{784}{p_{i}^{2}}\]
where $p_i$ is the pixel value ranging from $0$ to $255$ in row major.

Shallow quantum circuits are not only preferred in experiments due to a limited coherent time, but also preferred in numerical simulations for shorter computation time. To make the circuit as compact as possible while maintaining the fitting ability, ``Strongly entangled circuits'' \cite{schuld2020circuit}, composed of repeated $B_1$ and $B_3$ blocks as shown in Fig. \ref{fig:qnn-structure}, are adopted in our QNN structure. In each block $B_r$, all qubits are applied with $u3$ gates followed by $cu_3$ gates. For those control gates, the $i$-th qubit is selected as the control one and $(i+r)\mod n$ as the target. We choose the blocks $B_r$ with $r=1$ and $r=3$, so that all qubits are entangled in each block. To harness the fitting capacity of over parameterization (\cite{overpara}), we repeat the blocks for $\#$depth times. As shown in Fig. \ref{fig:qnn-structure}, each depth of circuit includes $40$ parameterized quantum gates ($u_3$ and $cu_3$) and $120$ trainable parameters recalling that $\omega_i$ is short for three parameters (see Tab. \ref{table:quantum-gate}).

Measurement operations are added at the end of the circuit to make predictions. The expectation values of the Pauli operator $\sigma_z$ for all qubits are measured, followed by a softmax function to generate a distribution over $10$ classes. Then, the cross entropy loss is calculated using the ground truth label and predicted probability distribution. Various optimization methods can be used to decrease the loss function, we adopt the Adam optimizer \cite{kingma2014adam} with these hyper-parameters: batch size $1024$, learning rate $\eta=0.01$, $\beta_1=0.9$, $\beta_2=0.999$.

\section{Quantum Teleportation}\label{Appendix-Qteleport}
Quantum teleportation\cite{QuantumTeleportation} can transmit a qubit in a long distance, which can be used to implement the quantum state transfer step in the QSA protocol proposed in this paper.\\
\begin{itemize}
    \item \textbf{First}, the server generates a pair of entangled qubits $\ket{\Phi^+}_{AB}=\frac{1}{\sqrt{2}}(\ket{0}_A\otimes\ket{0}_B+\ket{1}_A\otimes\ket{1}_B)$ and sends qubit B to one of the participant.
    \item \textbf{Second}, the participant receives qubit B. Meanwhile, he has a qubit C ($\ket{\Psi}_C = \alpha \ket{0}_C + \beta \ket{1}_C$) that prepared to transmit to the server. Now the system can be written as $\ket{\Psi}_C\otimes\ket{\Phi^+}_{AB} = \frac{1}{2}[
    \ket{\Phi^+}_{CB}\otimes(\alpha\ket{0}_A + \beta\ket{1}_A) + 
    \ket{\Phi^-}_{CB}\otimes(\alpha\ket{0}_A - \beta\ket{1}_A) + 
    \ket{\Psi^+}_{CB}\otimes(\alpha\ket{0}_A + \beta\ket{1}_A) + 
    \ket{\Psi^-}_{CB}\otimes(\alpha\ket{0}_A - \beta\ket{1}_A)])$, where the four Bell states are difined as $(\ket{\Phi^+} = \frac{1}{\sqrt{2}}(\ket{00}+\ket{11}), 
    \ket{\Phi^-} = \frac{1}{\sqrt{2}}(\ket{00}-\ket{11}), 
    \ket{\Psi^+} = \frac{1}{\sqrt{2}}(\ket{01}+\ket{10}), 
    \ket{\Psi^-} = \frac{1}{\sqrt{2}}(\ket{01}-\ket{10}))$.
    \item \textbf{Third}, the participant adapts $CNOT_{CB}$ and $H-gate$ on qubit C to entangle the three qubits. Then measures qubit C, B to get a $2$-bit information (one of $'00'$, $'01'$, $'10'$ and $'11'$, means $\ket{\Phi^+}$, $\ket{\Psi^+}$, $\ket{\Phi^-}$ and $\ket{\Psi^-}$ respectively) and sends the result to the server through \textit{classical channel}. Note that now the original qubit C is destroyed and the server has \textit{received} the qubit but has not reverted it to the original qubit C.
    \item \textbf{Fourth}, the server uses the classical message to decode the received qubit from the following strategy: $'00': I-gate(do \ nothing)$; $'01': X-gate$; $'10': Z-gate$; $'11': X-gate$ and $Z-gate$. Till now, the server receives the original qubit C.
\end{itemize}
\end{appendices}

\end{document}